\date{February 6, 2017}

\documentclass[12pt]{amsart}
\usepackage{latexsym,amsmath,amsfonts,amscd,amssymb}
\usepackage{graphicx}
\newtheorem{theorem*}{Theorem}

\theoremstyle{remark}

\title{Satoshi Risk Tables}


\author[C. Grunspan]{Cyril Grunspan}
\address{Cyril Grunspan\newline{}\indent L\'eonard de Vinci P\^ole Univ, Finance Lab\newline{}\indent Paris, France}
\email{cyril.grunspan@devinci.fr}

\author[R. P\'{e}rez-Marco]{Ricardo P\'{e}rez-Marco}
\address{Ricardo P\'{e}rez-Marco\newline{}\indent CNRS, IMJ-PRG, Labex R\'efi
, Labex MME-DDII\newline{}\indent Paris, France}
\email{ricardo.perez.marco@gmail.com}

\address{\tiny Author's Bitcoin Beer Address (ABBA)\footnote{\tiny Send some bitcoins to support our research at the pub.}:\newline{}\indent 1KrqVxqQFyUY9WuWcR5EHGVvhCS841LPLn} 

\address{\includegraphics[scale=0.5]{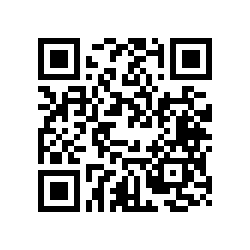}}

\begin{document}

\begin{abstract}
We present Bitcoin Security Tables computing the probability of success $p(z,q,t)$ of a double spend attack by an attacker controlling a share $q$ of the hashrate 
after $z$ confirmations in time $t$. 
\end{abstract}

\maketitle

\section{Introduction.}

The main breakthrough in \cite{N} is the solution to the \textit{double spend problem} of an electronic 
currency unit without a central authority. Bitcoin is the first 
form of \textit{peer-to-peer} (P2P) electronic currency. 

A double spend attack can only be attempted with a substantial fraction of the 
hashrate used in the \textit{Proof-of-Work} of the Bitcoin network. The attackers will start a \textit{double spend race}
against the rest of the network to replace the last blocks of the blockchain. 
The last section of \cite{N} computes the probability that the attackers catch up. Following Nakamoto, 
by ``success of the attackers'' we mean catching up the $z$ blocks, 
although to replace the blocks the attackers need to validate $z+1$. In \cite{GPM} we correct Nakamoto's analysis based on an abusive approximation  
and give a closed-form formula for the exact probability.

\begin{theorem*}\textnormal{(C. Grunspan, R. P\'erez-Marco, \cite{GPM})}

 Let $0<q<1/2$, resp. $p=1-q$, the relative hash power of the group of the attackers, resp. of honest miners. 
 After $z$ blocks have been validated by the honest miners, the probability of success of the attackers is 
 $$
 P(z)=I_{4pq}(z,1/2) \ ,
 $$
 where $I_x(a,b)$ is the Regularized Incomplete Beta Function
 $$
 I_x(a,b)=\frac{\Gamma(a+b)}{\Gamma(a) \Gamma(b)} \int_0^x t^{a-1}(1-t)^{b-1} \, dt \ .
 $$
\end{theorem*}

We carry out a more accurate risk analysis by considering not only the number of confirmations $z$ but also the time $t$ it took for the last $z$ 
validated blocks, which is an information that is clearly available.

In practice, in order to avoid a double spend attack, the recipient of the bitcoin transaction waits for  $z\geq 1$ 
confirmations. He also has the information on the time $t$ it took to confirm the transaction $z$ times. 
Obviously the probability of success of the attackers increases with $t$ since he has more time to secretly mine his alternative blockchain. 
In \cite{GPM} we carry out a more precise risk analysis considering this available data. The relevant dimensionless parameter introduced is the 
relative deviation from the expected time
$$
\kappa=\frac{t}{zt_0}=\frac{p t}{z\tau_0}=\frac{1-q}{z}\frac{t}{\tau_0} \ ,
$$
where $\tau_0$ is the expected validation time of a new block ($\tau_0=10 \, {\hbox{\rm min}}$ for the Bitcoin network).
In \cite{GPM} we give a closed-form formula for the probability $P(z,q, \kappa)$ that the attackers catch up with the current chain.

\begin{theorem*} \textnormal{(C. Grunspan, R. P\'erez-Marco, \cite{GPM})}

We have
$$
 P(z, q,  \kappa)=1-Q(z, \kappa z q/p) + \left (\frac{q}{p}\right )^{z} e^{\kappa z\frac{p-q}{p}} Q(z, \kappa z) \ .
$$
\end{theorem*}

After a validation $z$ has been observed in the network and the time $t$ being measured, and $r=\frac{t}{z t_0}$, one can compute the probabilities
$$
\tilde P(z, q, r) = P (z,q, (1-q) r) \ ,
$$
that give the probability of the attackers to catch up the current blockchain. We tabulate these probabilities for $z=1,2, \ldots , 10$.

\section{Use of the tables.}

The tables are useful to determine the number of confirmations needed for a given transaction. In practice, someone receiving a bitcoin transaction
will check after $z$ confirmations the corresponding table. With the measured time $t$, he will compute $r=\frac{t}{z t_0}$, 
and check the value of $\tilde P(z, q, r) $
and then asses the level of risk assumed accepting the payment.

\section{Satoshi Risk Tables.}

Below we give the tables for $z=1,2,\ldots, 9$ of $\tilde P(z, q, r)$ for different values of $q$ and $r$ in $\% $ with $2$ decimal places. These are what is 
needed for practical applications.

\pagebreak

\fbox{$$z=1$$}

$$
\begin{array}{|c||c|c|c|c|c|c|c|c|c|c|c|c|c|c|c|}
\hline
r \backslash q & 0.02 & 0.04 & 0.06 & 0.08 & 0.1 & 0.12 & 0.14 & 0.16 & 0.18 & 0.2 & 0.22 & 0.24 & 0.26\\ \hline \hline
0.1 & 2.24 & 4.55 & 6.94 & 9.42 & 12 & 14.67 & 17.44 & 20.33 & 23.34 & 26.49 & 29.77 & 33.2 & 36.8\\ \hline
0.2 & 2.43 & 4.93 & 7.5 & 10.14 & 12.87 & 15.68 & 18.59 & 21.6 & 24.71 & 27.94 & 31.3 & 34.79 & 38.42\\ \hline
0.3 & 2.63 & 5.31 & 8.05 & 10.86 & 13.74 & 16.69 & 19.72 & 22.84 & 26.05 & 29.37 & 32.79 & 36.33 & 40\\ \hline
0.4 & 2.82 & 5.69 & 8.6 & 11.57 & 14.6 & 17.68 & 20.84 & 24.07 & 27.37 & 30.77 & 34.25 & 37.84 & 41.54\\ \hline
0.5 & 3.02 & 6.06 & 9.15 & 12.28 & 15.45 & 18.67 & 21.94 & 25.27 & 28.67 & 32.14 & 35.68 & 39.32 & 43.04\\ \hline
0.6 & 3.21 & 6.44 & 9.69 & 12.97 & 16.29 & 19.64 & 23.02 & 26.46 & 29.94 & 33.48 & 37.08 & 40.76 & 44.5\\ \hline
0.7 & 3.4 & 6.81 & 10.23 & 13.67 & 17.12 & 20.59 & 24.09 & 27.62 & 31.19 & 34.8 & 38.45 & 42.16 & 45.93\\ \hline
0.8 & 3.6 & 7.18 & 10.77 & 14.36 & 17.95 & 21.54 & 25.15 & 28.77 & 32.42 & 36.09 & 39.79 & 43.53 & 47.32\\ \hline
0.9 & 3.79 & 7.56 & 11.3 & 15.04 & 18.76 & 22.48 & 26.19 & 29.9 & 33.62 & 37.35 & 41.1 & 44.87 & 48.67\\ \hline
1 & 3.98 & 7.92 & 11.83 & 15.72 & 19.57 & 23.4 & 27.22 & 31.02 & 34.81 & 38.6 & 42.38 & 46.18 & 49.99\\ \hline
1.1 & 4.17 & 8.29 & 12.36 & 16.39 & 20.37 & 24.32 & 28.23 & 32.11 & 35.97 & 39.81 & 43.64 & 47.45 & 51.27\\ \hline
1.2 & 4.36 & 8.66 & 12.89 & 17.05 & 21.16 & 25.22 & 29.23 & 33.19 & 37.11 & 41 & 44.86 & 48.7 & 52.52\\ \hline
1.3 & 4.55 & 9.02 & 13.41 & 17.71 & 21.95 & 26.11 & 30.21 & 34.25 & 38.24 & 42.17 & 46.06 & 49.92 & 53.74\\ \hline
1.4 & 4.75 & 9.39 & 13.93 & 18.37 & 22.72 & 26.99 & 31.18 & 35.29 & 39.34 & 43.32 & 47.24 & 51.1 & 54.93\\ \hline
1.5 & 4.94 & 9.75 & 14.44 & 19.02 & 23.49 & 27.86 & 32.14 & 36.32 & 40.42 & 44.44 & 48.38 & 52.26 & 56.08\\ \hline
1.6 & 5.13 & 10.11 & 14.95 & 19.67 & 24.25 & 28.72 & 33.08 & 37.33 & 41.48 & 45.54 & 49.51 & 53.4 & 57.21\\ \hline
1.7 & 5.32 & 10.47 & 15.46 & 20.31 & 25.01 & 29.57 & 34.01 & 38.33 & 42.53 & 46.62 & 50.61 & 54.5 & 58.31\\ \hline
1.8 & 5.5 & 10.82 & 15.97 & 20.94 & 25.75 & 30.41 & 34.93 & 39.31 & 43.55 & 47.67 & 51.68 & 55.58 & 59.38\\ \hline
1.9 & 5.69 & 11.18 & 16.47 & 21.57 & 26.49 & 31.24 & 35.83 & 40.27 & 44.56 & 48.71 & 52.73 & 56.63 & 60.42\\ \hline
2 & 5.88 & 11.53 & 16.97 & 22.2 & 27.22 & 32.06 & 36.73 & 41.22 & 45.55 & 49.73 & 53.76 & 57.66 & 61.44\\ \hline
2.1 & 6.07 & 11.89 & 17.47 & 22.82 & 27.95 & 32.87 & 37.6 & 42.15 & 46.52 & 50.72 & 54.77 & 58.67 & 62.43\\ \hline
2.2 & 6.26 & 12.24 & 17.96 & 23.43 & 28.66 & 33.68 & 38.47 & 43.07 & 47.47 & 51.7 & 55.75 & 59.65 & 63.39\\ \hline
2.3 & 6.44 & 12.59 & 18.45 & 24.04 & 29.37 & 34.47 & 39.33 & 43.97 & 48.41 & 52.65 & 56.71 & 60.6 & 64.33\\ \hline
2.4 & 6.63 & 12.94 & 18.94 & 24.65 & 30.08 & 35.25 & 40.17 & 44.86 & 49.33 & 53.59 & 57.66 & 61.54 & 65.25\\ \hline
2.5 & 6.82 & 13.29 & 19.42 & 25.25 & 30.77 & 36.02 & 41 & 45.74 & 50.23 & 54.51 & 58.58 & 62.45 & 66.14\\ \hline
2.6 & 7 & 13.63 & 19.91 & 25.84 & 31.46 & 36.78 & 41.82 & 46.6 & 51.12 & 55.41 & 59.48 & 63.34 & 67.01\\ \hline
2.7 & 7.19 & 13.98 & 20.38 & 26.43 & 32.14 & 37.54 & 42.63 & 47.44 & 51.99 & 56.29 & 60.36 & 64.21 & 67.85\\ \hline
2.8 & 7.38 & 14.32 & 20.86 & 27.02 & 32.82 & 38.28 & 43.43 & 48.28 & 52.85 & 57.16 & 61.22 & 65.06 & 68.68\\ \hline
2.9 & 7.56 & 14.66 & 21.33 & 27.6 & 33.49 & 39.02 & 44.22 & 49.1 & 53.69 & 58.01 & 62.07 & 65.89 & 69.48\\ \hline
3 & 7.75 & 15 & 21.8 & 28.18 & 34.15 & 39.75 & 44.99 & 49.91 & 54.52 & 58.84 & 62.89 & 66.7 & 70.27\\ \hline
3.1 & 7.93 & 15.34 & 22.27 & 28.75 & 34.8 & 40.46 & 45.76 & 50.7 & 55.33 & 59.65 & 63.7 & 67.49 & 71.03\\ \hline
3.2 & 8.11 & 15.68 & 22.74 & 29.32 & 35.45 & 41.18 & 46.51 & 51.49 & 56.13 & 60.45 & 64.49 & 68.26 & 71.77\\ \hline
3.3 & 8.3 & 16.02 & 23.2 & 29.88 & 36.1 & 41.88 & 47.25 & 52.26 & 56.91 & 61.24 & 65.26 & 69.01 & 72.5\\ \hline
3.4 & 8.48 & 16.35 & 23.66 & 30.44 & 36.73 & 42.57 & 47.99 & 53.01 & 57.68 & 62 & 66.02 & 69.74 & 73.2\\ \hline
3.5 & 8.66 & 16.69 & 24.12 & 30.99 & 37.36 & 43.26 & 48.71 & 53.76 & 58.43 & 62.76 & 66.76 & 70.46 & 73.89\\ 
\hline
\end{array}
$$

\pagebreak

\fbox{$z=2$}

$$
\begin{array}{|c||c|c|c|c|c|c|c|c|c|c|c|c|c|c|c|}
\hline
r \backslash q & 0.02 & 0.04 & 0.06 & 0.08 & 0.1 & 0.12 & 0.14 & 0.16 & 0.18 & 0.2 & 0.22 & 0.24 & 0.26\\ \hline \hline
0.1 & 0.05 & 0.21 & 0.49 & 0.89 & 1.45 & 2.16 & 3.06 & 4.15 & 5.47 & 7.04 & 8.89 & 11.06 & 13.58\\ \hline
0.2 & 0.06 & 0.25 & 0.58 & 1.05 & 1.69 & 2.51 & 3.52 & 4.74 & 6.2 & 7.92 & 9.92 & 12.25 & 14.92\\ \hline
0.3 & 0.07 & 0.3 & 0.68 & 1.23 & 1.96 & 2.89 & 4.03 & 5.39 & 7 & 8.87 & 11.03 & 13.51 & 16.35\\ \hline
0.4 & 0.09 & 0.35 & 0.79 & 1.43 & 2.26 & 3.31 & 4.58 & 6.09 & 7.85 & 9.89 & 12.21 & 14.86 & 17.85\\ \hline
0.5 & 0.1 & 0.4 & 0.92 & 1.64 & 2.59 & 3.77 & 5.18 & 6.84 & 8.76 & 10.96 & 13.46 & 16.26 & 19.41\\ \hline
0.6 & 0.12 & 0.47 & 1.05 & 1.88 & 2.94 & 4.25 & 5.81 & 7.64 & 9.73 & 12.09 & 14.76 & 17.73 & 21.02\\ \hline
0.7 & 0.13 & 0.54 & 1.2 & 2.13 & 3.32 & 4.77 & 6.49 & 8.47 & 10.73 & 13.27 & 16.1 & 19.24 & 22.68\\ \hline
0.8 & 0.15 & 0.61 & 1.36 & 2.4 & 3.72 & 5.32 & 7.2 & 9.35 & 11.78 & 14.5 & 17.49 & 20.78 & 24.37\\ \hline
0.9 & 0.17 & 0.69 & 1.53 & 2.68 & 4.14 & 5.89 & 7.94 & 10.26 & 12.87 & 15.76 & 18.92 & 22.36 & 26.09\\ \hline
1 & 0.2 & 0.77 & 1.71 & 2.98 & 4.58 & 6.5 & 8.71 & 11.21 & 13.99 & 17.05 & 20.37 & 23.97 & 27.83\\ \hline
1.1 & 0.22 & 0.86 & 1.89 & 3.3 & 5.05 & 7.12 & 9.51 & 12.18 & 15.14 & 18.37 & 21.86 & 25.6 & 29.59\\ \hline
1.2 & 0.24 & 0.95 & 2.09 & 3.63 & 5.53 & 7.77 & 10.33 & 13.18 & 16.32 & 19.71 & 23.36 & 27.24 & 31.35\\ \hline
1.3 & 0.27 & 1.05 & 2.3 & 3.97 & 6.03 & 8.44 & 11.18 & 14.21 & 17.52 & 21.08 & 24.87 & 28.89 & 33.11\\ \hline
1.4 & 0.3 & 1.15 & 2.51 & 4.33 & 6.54 & 9.13 & 12.04 & 15.26 & 18.74 & 22.46 & 26.4 & 30.54 & 34.87\\ \hline
1.5 & 0.33 & 1.26 & 2.74 & 4.69 & 7.08 & 9.84 & 12.93 & 16.32 & 19.97 & 23.85 & 27.94 & 32.2 & 36.63\\ \hline
1.6 & 0.36 & 1.37 & 2.97 & 5.08 & 7.63 & 10.56 & 13.84 & 17.41 & 21.22 & 25.26 & 29.48 & 33.85 & 38.37\\ \hline
1.7 & 0.39 & 1.49 & 3.21 & 5.47 & 8.19 & 11.31 & 14.76 & 18.5 & 22.48 & 26.67 & 31.02 & 35.5 & 40.1\\ \hline
1.8 & 0.42 & 1.61 & 3.46 & 5.88 & 8.77 & 12.06 & 15.7 & 19.61 & 23.76 & 28.08 & 32.55 & 37.14 & 41.81\\ \hline
1.9 & 0.46 & 1.74 & 3.72 & 6.29 & 9.36 & 12.84 & 16.65 & 20.73 & 25.03 & 29.5 & 34.09 & 38.77 & 43.5\\ \hline
2 & 0.49 & 1.87 & 3.98 & 6.72 & 9.96 & 13.62 & 17.61 & 21.87 & 26.32 & 30.92 & 35.62 & 40.38 & 45.17\\ \hline
2.1 & 0.53 & 2 & 4.26 & 7.15 & 10.58 & 14.42 & 18.59 & 23 & 27.6 & 32.33 & 37.13 & 41.97 & 46.82\\ \hline
2.2 & 0.57 & 2.14 & 4.54 & 7.6 & 11.2 & 15.22 & 19.57 & 24.15 & 28.89 & 33.74 & 38.64 & 43.55 & 48.44\\ \hline
2.3 & 0.61 & 2.28 & 4.82 & 8.06 & 11.84 & 16.04 & 20.56 & 25.29 & 30.18 & 35.14 & 40.13 & 45.11 & 50.03\\ \hline
2.4 & 0.65 & 2.43 & 5.11 & 8.52 & 12.48 & 16.87 & 21.55 & 26.45 & 31.46 & 36.54 & 41.61 & 46.64 & 51.6\\ \hline
2.5 & 0.69 & 2.58 & 5.41 & 8.99 & 13.14 & 17.7 & 22.56 & 27.6 & 32.74 & 37.92 & 43.07 & 48.15 & 53.13\\ \hline
2.6 & 0.73 & 2.73 & 5.72 & 9.47 & 13.8 & 18.54 & 23.56 & 28.75 & 34.02 & 39.29 & 44.52 & 49.64 & 54.63\\ \hline
2.7 & 0.78 & 2.89 & 6.03 & 9.96 & 14.47 & 19.39 & 24.57 & 29.9 & 35.29 & 40.66 & 45.94 & 51.1 & 56.1\\ \hline
2.8 & 0.82 & 3.05 & 6.35 & 10.46 & 15.15 & 20.24 & 25.58 & 31.05 & 36.55 & 42 & 47.35 & 52.54 & 57.54\\ \hline
2.9 & 0.87 & 3.21 & 6.67 & 10.96 & 15.84 & 21.1 & 26.6 & 32.2 & 37.81 & 43.34 & 48.73 & 53.95 & 58.95\\ \hline
3 & 0.92 & 3.38 & 7 & 11.47 & 16.53 & 21.96 & 27.61 & 33.34 & 39.05 & 44.66 & 50.1 & 55.33 & 60.32\\ \hline
3.1 & 0.97 & 3.55 & 7.34 & 11.99 & 17.22 & 22.83 & 28.63 & 34.48 & 40.29 & 45.96 & 51.44 & 56.68 & 61.65\\ \hline
3.2 & 1.02 & 3.73 & 7.68 & 12.51 & 17.92 & 23.7 & 29.64 & 35.62 & 41.51 & 47.24 & 52.75 & 58 & 62.96\\ \hline
3.3 & 1.07 & 3.91 & 8.02 & 13.03 & 18.63 & 24.57 & 30.65 & 36.74 & 42.72 & 48.51 & 54.05 & 59.3 & 64.23\\ \hline
3.4 & 1.12 & 4.09 & 8.37 & 13.57 & 19.34 & 25.44 & 31.66 & 37.86 & 43.92 & 49.76 & 55.32 & 60.56 & 65.47\\ \hline
3.5 & 1.18 & 4.27 & 8.73 & 14.1 & 20.06 & 26.31 & 32.67 & 38.97 & 45.11 & 50.99 & 56.56 & 61.8 & 66.67\\ 
\hline
\end{array}
$$

\pagebreak

\fbox{$z=3$}

$$
\begin{array}{|c||c|c|c|c|c|c|c|c|c|c|c|c|c|c|c|}
\hline
r \backslash q & 0.02 & 0.04 & 0.06 & 0.08 & 0.1 & 0.12 & 0.14 & 0.16 & 0.18 & 0.2 & 0.22 & 0.24 & 0.26\\ \hline \hline
0.1 & 0 & 0.01 & 0.03 & 0.08 & 0.17 & 0.32 & 0.54 & 0.85 & 1.28 & 1.87 & 2.65 & 3.68 & 5.01\\ \hline
0.2 & 0 & 0.01 & 0.04 & 0.11 & 0.22 & 0.4 & 0.66 & 1.04 & 1.55 & 2.24 & 3.14 & 4.3 & 5.78\\ \hline
0.3 & 0 & 0.02 & 0.06 & 0.14 & 0.28 & 0.5 & 0.82 & 1.27 & 1.87 & 2.67 & 3.7 & 5.01 & 6.66\\ \hline
0.4 & 0 & 0.02 & 0.07 & 0.18 & 0.35 & 0.62 & 1 & 1.54 & 2.25 & 3.17 & 4.34 & 5.82 & 7.65\\ \hline
0.5 & 0 & 0.03 & 0.09 & 0.22 & 0.44 & 0.76 & 1.22 & 1.85 & 2.68 & 3.74 & 5.07 & 6.72 & 8.75\\ \hline
0.6 & 0 & 0.03 & 0.12 & 0.27 & 0.54 & 0.93 & 1.48 & 2.22 & 3.17 & 4.39 & 5.89 & 7.73 & 9.95\\ \hline
0.7 & 0.01 & 0.04 & 0.14 & 0.34 & 0.65 & 1.12 & 1.77 & 2.63 & 3.73 & 5.1 & 6.79 & 8.83 & 11.26\\ \hline
0.8 & 0.01 & 0.05 & 0.18 & 0.41 & 0.79 & 1.34 & 2.09 & 3.09 & 4.34 & 5.9 & 7.78 & 10.03 & 12.67\\ \hline
0.9 & 0.01 & 0.06 & 0.21 & 0.49 & 0.94 & 1.58 & 2.46 & 3.6 & 5.02 & 6.76 & 8.85 & 11.31 & 14.17\\ \hline
1 & 0.01 & 0.08 & 0.25 & 0.58 & 1.11 & 1.86 & 2.86 & 4.15 & 5.76 & 7.7 & 9.99 & 12.67 & 15.76\\ \hline
1.1 & 0.01 & 0.09 & 0.3 & 0.69 & 1.3 & 2.16 & 3.3 & 4.76 & 6.56 & 8.7 & 11.21 & 14.12 & 17.43\\ \hline
1.2 & 0.01 & 0.11 & 0.35 & 0.8 & 1.5 & 2.49 & 3.79 & 5.42 & 7.41 & 9.77 & 12.5 & 15.63 & 19.16\\ \hline
1.3 & 0.02 & 0.13 & 0.41 & 0.93 & 1.73 & 2.85 & 4.31 & 6.13 & 8.32 & 10.9 & 13.86 & 17.21 & 20.96\\ \hline
1.4 & 0.02 & 0.15 & 0.48 & 1.07 & 1.98 & 3.23 & 4.86 & 6.88 & 9.28 & 12.08 & 15.27 & 18.85 & 22.81\\ \hline
1.5 & 0.02 & 0.17 & 0.55 & 1.22 & 2.25 & 3.65 & 5.46 & 7.67 & 10.3 & 13.32 & 16.74 & 20.54 & 24.7\\ \hline
1.6 & 0.03 & 0.2 & 0.63 & 1.39 & 2.54 & 4.1 & 6.09 & 8.51 & 11.36 & 14.62 & 18.26 & 22.27 & 26.63\\ \hline
1.7 & 0.03 & 0.23 & 0.71 & 1.57 & 2.84 & 4.57 & 6.76 & 9.39 & 12.47 & 15.95 & 19.82 & 24.05 & 28.59\\ \hline
1.8 & 0.03 & 0.26 & 0.8 & 1.76 & 3.17 & 5.07 & 7.46 & 10.31 & 13.62 & 17.33 & 21.42 & 25.85 & 30.57\\ \hline
1.9 & 0.04 & 0.29 & 0.9 & 1.96 & 3.52 & 5.6 & 8.19 & 11.27 & 14.81 & 18.75 & 23.06 & 27.68 & 32.57\\ \hline
2 & 0.04 & 0.33 & 1 & 2.18 & 3.89 & 6.16 & 8.96 & 12.27 & 16.03 & 20.2 & 24.72 & 29.53 & 34.58\\ \hline
2.1 & 0.05 & 0.36 & 1.12 & 2.41 & 4.28 & 6.74 & 9.76 & 13.3 & 17.29 & 21.68 & 26.41 & 31.39 & 36.59\\ \hline
2.2 & 0.06 & 0.4 & 1.23 & 2.65 & 4.69 & 7.35 & 10.59 & 14.36 & 18.58 & 23.19 & 28.11 & 33.27 & 38.59\\ \hline
2.3 & 0.06 & 0.45 & 1.36 & 2.9 & 5.12 & 7.98 & 11.44 & 15.44 & 19.9 & 24.72 & 29.83 & 35.14 & 40.58\\ \hline
2.4 & 0.07 & 0.49 & 1.49 & 3.17 & 5.56 & 8.64 & 12.33 & 16.56 & 21.24 & 26.27 & 31.56 & 37.01 & 42.56\\ \hline
2.5 & 0.08 & 0.54 & 1.63 & 3.46 & 6.03 & 9.31 & 13.24 & 17.7 & 22.6 & 27.83 & 33.29 & 38.88 & 44.53\\ \hline
2.6 & 0.08 & 0.59 & 1.78 & 3.75 & 6.51 & 10.02 & 14.17 & 18.87 & 23.98 & 29.41 & 35.02 & 40.74 & 46.46\\ \hline
2.7 & 0.09 & 0.65 & 1.94 & 4.06 & 7.01 & 10.74 & 15.13 & 20.05 & 25.38 & 30.99 & 36.76 & 42.58 & 48.38\\ \hline
2.8 & 0.1 & 0.71 & 2.1 & 4.38 & 7.53 & 11.48 & 16.1 & 21.25 & 26.79 & 32.57 & 38.48 & 44.41 & 50.26\\ \hline
2.9 & 0.11 & 0.77 & 2.27 & 4.71 & 8.07 & 12.25 & 17.1 & 22.47 & 28.21 & 34.16 & 40.2 & 46.22 & 52.11\\ \hline
3 & 0.12 & 0.83 & 2.45 & 5.06 & 8.62 & 13.03 & 18.11 & 23.71 & 29.64 & 35.75 & 41.91 & 48 & 53.92\\ \hline
3.1 & 0.13 & 0.9 & 2.63 & 5.41 & 9.19 & 13.83 & 19.14 & 24.95 & 31.07 & 37.33 & 43.6 & 49.75 & 55.7\\ \hline
3.2 & 0.14 & 0.97 & 2.82 & 5.78 & 9.78 & 14.64 & 20.19 & 26.21 & 32.51 & 38.91 & 45.28 & 51.48 & 57.43\\ \hline
3.3 & 0.15 & 1.04 & 3.02 & 6.16 & 10.38 & 15.48 & 21.25 & 27.47 & 33.95 & 40.48 & 46.93 & 53.18 & 59.13\\ \hline
3.4 & 0.16 & 1.12 & 3.23 & 6.56 & 10.99 & 16.32 & 22.32 & 28.75 & 35.38 & 42.04 & 48.57 & 54.84 & 60.78\\ \hline
3.5 & 0.18 & 1.2 & 3.44 & 6.96 & 11.62 & 17.18 & 23.41 & 30.03 & 36.82 & 43.59 & 50.18 & 56.47 & 62.39\\ 
\hline
\end{array}
$$

\pagebreak

\fbox{$z=4$}

$$
\begin{array}{|c||c|c|c|c|c|c|c|c|c|c|c|c|c|c|c|}
\hline
r \backslash q & 0.02 & 0.04 & 0.06 & 0.08 & 0.1 & 0.12 & 0.14 & 0.16 & 0.18 & 0.2 & 0.22 & 0.24 & 0.26\\ \hline \hline
0.1 & 0 & 0 & 0 & 0.01 & 0.02 & 0.05 & 0.09 & 0.17 & 0.3 & 0.5 & 0.79 & 1.22 & 1.85\\ \hline 
0.2 & 0 & 0 & 0 & 0.01 & 0.03 & 0.06 & 0.12 & 0.23 & 0.39 & 0.63 & 0.99 & 1.51 & 2.24\\ \hline 
0.3 & 0 & 0 & 0 & 0.02 & 0.04 & 0.09 & 0.17 & 0.3 & 0.5 & 0.8 & 1.24 & 1.85 & 2.71\\ \hline 
0.4 & 0 & 0 & 0.01 & 0.02 & 0.05 & 0.12 & 0.22 & 0.39 & 0.64 & 1.01 & 1.54 & 2.27 & 3.27\\ \hline 
0.5 & 0 & 0 & 0.01 & 0.03 & 0.07 & 0.15 & 0.29 & 0.5 & 0.82 & 1.27 & 1.91 & 2.77 & 3.93\\ \hline 
0.6 & 0 & 0 & 0.01 & 0.04 & 0.1 & 0.2 & 0.37 & 0.64 & 1.03 & 1.59 & 2.35 & 3.36 & 4.7\\ \hline 
0.7 & 0 & 0 & 0.02 & 0.05 & 0.13 & 0.26 & 0.48 & 0.81 & 1.3 & 1.96 & 2.87 & 4.05 & 5.59\\ \hline 
0.8 & 0 & 0 & 0.02 & 0.07 & 0.17 & 0.34 & 0.61 & 1.02 & 1.61 & 2.41 & 3.47 & 4.84 & 6.6\\ \hline 
0.9 & 0 & 0.01 & 0.03 & 0.09 & 0.21 & 0.43 & 0.77 & 1.27 & 1.97 & 2.92 & 4.16 & 5.74 & 7.73\\ \hline 
1 & 0 & 0.01 & 0.04 & 0.12 & 0.27 & 0.54 & 0.95 & 1.55 & 2.39 & 3.5 & 4.94 & 6.74 & 8.97\\ \hline 
1.1 & 0 & 0.01 & 0.05 & 0.15 & 0.34 & 0.66 & 1.16 & 1.89 & 2.87 & 4.16 & 5.81 & 7.85 & 10.34\\ \hline
1.2 & 0 & 0.01 & 0.06 & 0.18 & 0.42 & 0.81 & 1.41 & 2.26 & 3.41 & 4.9 & 6.77 & 9.07 & 11.83\\ \hline 
1.3 & 0 & 0.02 & 0.08 & 0.22 & 0.51 & 0.98 & 1.69 & 2.69 & 4.02 & 5.72 & 7.83 & 10.39 & 13.42\\ \hline 
1.4 & 0 & 0.02 & 0.09 & 0.27 & 0.61 & 1.17 & 2 & 3.16 & 4.68 & 6.61 & 8.98 & 11.81 & 15.12\\ \hline 
1.5 & 0 & 0.02 & 0.11 & 0.33 & 0.73 & 1.39 & 2.35 & 3.68 & 5.42 & 7.58 & 10.21 & 13.32 & 16.91\\ \hline 
1.6 & 0 & 0.03 & 0.14 & 0.39 & 0.86 & 1.63 & 2.74 & 4.26 & 6.21 & 8.63 & 11.53 & 14.92 & 18.79\\ \hline 
1.7 & 0 & 0.04 & 0.16 & 0.46 & 1.01 & 1.9 & 3.17 & 4.88 & 7.07 & 9.75 & 12.93 & 16.6 & 20.75\\ \hline 
1.8 & 0 & 0.04 & 0.19 & 0.54 & 1.18 & 2.19 & 3.64 & 5.56 & 7.99 & 10.94 & 14.4 & 18.35 & 22.77\\ \hline 
1.9 & 0 & 0.05 & 0.22 & 0.63 & 1.37 & 2.52 & 4.14 & 6.29 & 8.98 & 12.2 & 15.95 & 20.18 & 24.86\\ \hline 
2 & 0 & 0.06 & 0.26 & 0.73 & 1.57 & 2.87 & 4.69 & 7.07 & 10.02 & 13.53 & 17.55 & 22.06 & 27\\ \hline 
2.1 & 0 & 0.07 & 0.3 & 0.84 & 1.79 & 3.25 & 5.28 & 7.9 & 11.12 & 14.91 & 19.22 & 24 & 29.17\\ \hline 
2.2 & 0.01 & 0.08 & 0.35 & 0.96 & 2.03 & 3.66 & 5.9 & 8.78 & 12.28 & 16.35 & 20.94 & 25.98 & 31.38\\ \hline
2.3 & 0.01 & 0.09 & 0.4 & 1.08 & 2.29 & 4.1 & 6.57 & 9.71 & 13.48 & 17.84 & 22.71 & 28 & 33.61\\ \hline 
2.4 & 0.01 & 0.1 & 0.45 & 1.22 & 2.56 & 4.57 & 7.27 & 10.68 & 14.74 & 19.38 & 24.52 & 30.04 & 35.85\\ \hline
2.5 & 0.01 & 0.12 & 0.51 & 1.38 & 2.86 & 5.07 & 8.02 & 11.69 & 16.04 & 20.97 & 26.36 & 32.11 & 38.09\\ \hline 
2.6 & 0.01 & 0.13 & 0.58 & 1.54 & 3.18 & 5.59 & 8.79 & 12.75 & 17.39 & 22.59 & 28.23 & 34.19 & 40.34\\ \hline 
2.7 & 0.01 & 0.15 & 0.65 & 1.71 & 3.52 & 6.15 & 9.61 & 13.85 & 18.77 & 24.24 & 30.13 & 36.28 & 42.57\\ \hline 
2.8 & 0.01 & 0.17 & 0.72 & 1.9 & 3.88 & 6.74 & 10.46 & 14.99 & 20.19 & 25.93 & 32.04 & 38.37 & 44.78\\ \hline 
2.9 & 0.01 & 0.19 & 0.8 & 2.1 & 4.26 & 7.35 & 11.35 & 16.16 & 21.64 & 27.63 & 33.96 & 40.46 & 46.97\\ \hline 
3 & 0.02 & 0.21 & 0.89 & 2.31 & 4.66 & 7.99 & 12.26 & 17.36 & 23.12 & 29.36 & 35.89 & 42.53 & 49.13\\ \hline 
3.1 & 0.02 & 0.24 & 0.98 & 2.54 & 5.08 & 8.66 & 13.21 & 18.6 & 24.63 & 31.1 & 37.82 & 44.59 & 51.26\\ \hline 
3.2 & 0.02 & 0.26 & 1.08 & 2.78 & 5.53 & 9.36 & 14.19 & 19.86 & 26.16 & 32.86 & 39.75 & 46.63 & 53.35\\ \hline 
3.3 & 0.02 & 0.29 & 1.18 & 3.03 & 5.99 & 10.08 & 15.2 & 21.15 & 27.71 & 34.62 & 41.67 & 48.65 & 55.39\\ \hline 
3.4 & 0.02 & 0.32 & 1.3 & 3.29 & 6.47 & 10.83 & 16.24 & 22.47 & 29.27 & 36.38 & 43.57 & 50.63 & 57.39\\ \hline 
3.5 & 0.03 & 0.35 & 1.41 & 3.57 & 6.98 & 11.61 & 17.3 & 23.8 & 30.85 & 38.15 & 45.46 & 52.58 & 59.34\\ 
\hline
\end{array}
$$

\pagebreak

\fbox{$z=5$}

$$
\begin{array}{|c||c|c|c|c|c|c|c|c|c|c|c|c|c|c|c|}
\hline
r \backslash q & 0.02 & 0.04 & 0.06 & 0.08 & 0.1 & 0.12 & 0.14 & 0.16 & 0.18 & 0.2 & 0.22 & 0.24 & 0.26\\ \hline \hline
0.1 & 0 & 0 & 0 & 0 & 0 & 0.01 & 0.02 & 0.04 & 0.07 & 0.13 & 0.24 & 0.41 & 0.68\\ \hline 
0.2 & 0 & 0 & 0 & 0 & 0 & 0.01 & 0.02 & 0.05 & 0.1 & 0.18 & 0.31 & 0.53 & 0.87\\ \hline 
0.3 & 0 & 0 & 0 & 0 & 0.01 & 0.01 & 0.03 & 0.07 & 0.13 & 0.24 & 0.41 & 0.68 & 1.1\\ \hline 
0.4 & 0 & 0 & 0 & 0 & 0.01 & 0.02 & 0.05 & 0.1 & 0.18 & 0.32 & 0.54 & 0.89 & 1.39\\ \hline 
0.5 & 0 & 0 & 0 & 0 & 0.01 & 0.03 & 0.07 & 0.13 & 0.25 & 0.43 & 0.72 & 1.14 & 1.76\\ \hline 
0.6 & 0 & 0 & 0 & 0.01 & 0.02 & 0.04 & 0.09 & 0.19 & 0.34 & 0.57 & 0.93 & 1.46 & 2.22\\ \hline 
0.7 & 0 & 0 & 0 & 0.01 & 0.03 & 0.06 & 0.13 & 0.25 & 0.45 & 0.75 & 1.21 & 1.86 & 2.77\\ \hline 
0.8 & 0 & 0 & 0 & 0.01 & 0.04 & 0.09 & 0.18 & 0.34 & 0.59 & 0.98 & 1.55 & 2.34 & 3.43\\ \hline 
0.9 & 0 & 0 & 0 & 0.02 & 0.05 & 0.12 & 0.24 & 0.45 & 0.77 & 1.26 & 1.96 & 2.92 & 4.21\\ \hline 
1 & 0 & 0 & 0.01 & 0.02 & 0.07 & 0.16 & 0.32 & 0.58 & 1 & 1.6 & 2.45 & 3.6 & 5.12\\ \hline 
1.1 & 0 & 0 & 0.01 & 0.03 & 0.09 & 0.21 & 0.41 & 0.75 & 1.26 & 2 & 3.02 & 4.39 & 6.16\\ \hline 
1.2 & 0 & 0 & 0.01 & 0.04 & 0.12 & 0.27 & 0.53 & 0.95 & 1.58 & 2.48 & 3.69 & 5.29 & 7.34\\ \hline 
1.3 & 0 & 0 & 0.01 & 0.05 & 0.15 & 0.34 & 0.67 & 1.19 & 1.96 & 3.03 & 4.46 & 6.31 & 8.65\\ \hline 
1.4 & 0 & 0 & 0.02 & 0.07 & 0.19 & 0.43 & 0.83 & 1.47 & 2.39 & 3.65 & 5.32 & 7.46 & 10.1\\ \hline 
1.5 & 0 & 0 & 0.02 & 0.09 & 0.24 & 0.53 & 1.03 & 1.79 & 2.88 & 4.36 & 6.29 & 8.72 & 11.68\\ \hline 
1.6 & 0 & 0 & 0.03 & 0.11 & 0.3 & 0.66 & 1.25 & 2.16 & 3.44 & 5.16 & 7.36 & 10.1 & 13.39\\ \hline 
1.7 & 0 & 0.01 & 0.04 & 0.14 & 0.37 & 0.8 & 1.51 & 2.58 & 4.07 & 6.04 & 8.53 & 11.59 & 15.22\\ \hline 
1.8 & 0 & 0.01 & 0.05 & 0.17 & 0.45 & 0.96 & 1.8 & 3.05 & 4.76 & 7 & 9.81 & 13.19 & 17.16\\ \hline 
1.9 & 0 & 0.01 & 0.06 & 0.21 & 0.54 & 1.15 & 2.13 & 3.57 & 5.53 & 8.05 & 11.18 & 14.9 & 19.2\\ \hline 
2 & 0 & 0.01 & 0.07 & 0.25 & 0.64 & 1.36 & 2.5 & 4.14 & 6.36 & 9.19 & 12.64 & 16.7 & 21.34\\ \hline 
2.1 & 0 & 0.01 & 0.08 & 0.3 & 0.76 & 1.59 & 2.9 & 4.78 & 7.27 & 10.41 & 14.19 & 18.59 & 23.55\\ \hline 
2.2 & 0 & 0.02 & 0.1 & 0.35 & 0.89 & 1.86 & 3.35 & 5.46 & 8.25 & 11.71 & 15.83 & 20.57 & 25.84\\ \hline 
2.3 & 0 & 0.02 & 0.12 & 0.41 & 1.04 & 2.15 & 3.84 & 6.21 & 9.29 & 13.08 & 17.55 & 22.61 & 28.19\\ \hline 
2.4 & 0 & 0.02 & 0.14 & 0.48 & 1.21 & 2.46 & 4.37 & 7.01 & 10.4 & 14.53 & 19.33 & 24.72 & 30.58\\ \hline 
2.5 & 0 & 0.03 & 0.16 & 0.56 & 1.39 & 2.81 & 4.95 & 7.86 & 11.58 & 16.05 & 21.19 & 26.88 & 33.01\\ \hline 
2.6 & 0 & 0.03 & 0.19 & 0.65 & 1.59 & 3.19 & 5.56 & 8.78 & 12.82 & 17.63 & 23.1 & 29.09 & 35.46\\ \hline 
2.7 & 0 & 0.04 & 0.22 & 0.74 & 1.81 & 3.59 & 6.22 & 9.74 & 14.12 & 19.27 & 25.06 & 31.33 & 37.92\\ \hline 
2.8 & 0 & 0.04 & 0.25 & 0.84 & 2.04 & 4.03 & 6.93 & 10.76 & 15.47 & 20.96 & 27.06 & 33.6 & 40.39\\ \hline 
2.9 & 0 & 0.05 & 0.29 & 0.96 & 2.3 & 4.5 & 7.68 & 11.83 & 16.88 & 22.7 & 29.1 & 35.89 & 42.85\\ \hline 
3 & 0 & 0.06 & 0.33 & 1.08 & 2.58 & 5 & 8.46 & 12.95 & 18.34 & 24.49 & 31.17 & 38.18 & 45.29\\ \hline 
3.1 & 0 & 0.06 & 0.37 & 1.22 & 2.87 & 5.54 & 9.3 & 14.11 & 19.85 & 26.31 & 33.26 & 40.47 & 47.7\\ \hline 
3.2 & 0 & 0.07 & 0.42 & 1.36 & 3.19 & 6.1 & 10.17 & 15.32 & 21.39 & 28.16 & 35.37 & 42.75 & 50.09\\ \hline 
3.3 & 0 & 0.08 & 0.48 & 1.52 & 3.53 & 6.7 & 11.08 & 16.57 & 22.98 & 30.04 & 37.48 & 45.02 & 52.43\\ \hline 
3.4 & 0 & 0.09 & 0.53 & 1.69 & 3.89 & 7.33 & 12.03 & 17.87 & 24.6 & 31.94 & 39.59 & 47.27 & 54.72\\ \hline 
3.5 & 0 & 0.11 & 0.59 & 1.87 & 4.28 & 8 & 13.02 & 19.19 & 26.25 & 33.86 & 41.7 & 49.49 & 56.97\\ 
\hline
\end{array}
$$

\pagebreak

\fbox{$z=6$}

$$
\begin{array}{|c||c|c|c|c|c|c|c|c|c|c|c|c|c|c|c|}
\hline
r \backslash q & 0.02 & 0.04 & 0.06 & 0.08 & 0.1 & 0.12 & 0.14 & 0.16 & 0.18 & 0.2 & 0.22 & 0.24 & 0.26\\ \hline \hline
0.1 & 0 & 0 & 0 & 0 & 0 & 0 & 0 & 0.01 & 0.02 & 0.03 & 0.07 & 0.14 & 0.25\\ \hline 
0.2 & 0 & 0 & 0 & 0 & 0 & 0 & 0 & 0.01 & 0.02 & 0.05 & 0.1 & 0.19 & 0.33\\ \hline 
0.3 & 0 & 0 & 0 & 0 & 0 & 0 & 0.01 & 0.02 & 0.04 & 0.07 & 0.14 & 0.25 & 0.45\\ \hline 
0.4 & 0 & 0 & 0 & 0 & 0 & 0 & 0.01 & 0.02 & 0.05 & 0.1 & 0.19 & 0.34 & 0.59\\ \hline 
0.5 & 0 & 0 & 0 & 0 & 0 & 0.01 & 0.02 & 0.04 & 0.08 & 0.15 & 0.27 & 0.47 & 0.79\\ \hline 
0.6 & 0 & 0 & 0 & 0 & 0 & 0.01 & 0.02 & 0.05 & 0.11 & 0.21 & 0.37 & 0.63 & 1.04\\ \hline 
0.7 & 0 & 0 & 0 & 0 & 0 & 0.01 & 0.04 & 0.08 & 0.16 & 0.29 & 0.51 & 0.85 & 1.37\\ \hline 
0.8 & 0 & 0 & 0 & 0 & 0.01 & 0.02 & 0.05 & 0.11 & 0.22 & 0.4 & 0.69 & 1.13 & 1.78\\ \hline 
0.9 & 0 & 0 & 0 & 0 & 0.01 & 0.03 & 0.08 & 0.16 & 0.3 & 0.55 & 0.92 & 1.48 & 2.3\\ \hline 
1 & 0 & 0 & 0 & 0 & 0.02 & 0.05 & 0.11 & 0.22 & 0.42 & 0.73 & 1.21 & 1.92 & 2.93\\ \hline 
1.1 & 0 & 0 & 0 & 0.01 & 0.02 & 0.06 & 0.15 & 0.3 & 0.56 & 0.97 & 1.58 & 2.46 & 3.68\\ \hline 
1.2 & 0 & 0 & 0 & 0.01 & 0.03 & 0.09 & 0.2 & 0.4 & 0.74 & 1.26 & 2.02 & 3.1 & 4.57\\ \hline 
1.3 & 0 & 0 & 0 & 0.01 & 0.04 & 0.12 & 0.27 & 0.53 & 0.96 & 1.61 & 2.55 & 3.86 & 5.6\\ \hline 
1.4 & 0 & 0 & 0 & 0.02 & 0.06 & 0.16 & 0.35 & 0.69 & 1.23 & 2.03 & 3.18 & 4.74 & 6.78\\ \hline 
1.5 & 0 & 0 & 0.01 & 0.02 & 0.08 & 0.21 & 0.45 & 0.88 & 1.54 & 2.53 & 3.9 & 5.74 & 8.11\\ \hline 
1.6 & 0 & 0 & 0.01 & 0.03 & 0.1 & 0.27 & 0.58 & 1.1 & 1.92 & 3.11 & 4.74 & 6.88 & 9.6\\ \hline 
1.7 & 0 & 0 & 0.01 & 0.04 & 0.14 & 0.34 & 0.73 & 1.37 & 2.36 & 3.77 & 5.68 & 8.15 & 11.24\\ \hline 
1.8 & 0 & 0 & 0.01 & 0.05 & 0.17 & 0.43 & 0.9 & 1.69 & 2.86 & 4.52 & 6.74 & 9.56 & 13.02\\ \hline 
1.9 & 0 & 0 & 0.01 & 0.07 & 0.22 & 0.53 & 1.11 & 2.05 & 3.44 & 5.37 & 7.91 & 11.09 & 14.95\\ \hline 
2 & 0 & 0 & 0.02 & 0.09 & 0.27 & 0.65 & 1.35 & 2.46 & 4.08 & 6.31 & 9.19 & 12.75 & 17\\ \hline 
2.1 & 0 & 0 & 0.02 & 0.11 & 0.33 & 0.79 & 1.62 & 2.92 & 4.8 & 7.34 & 10.58 & 14.54 & 19.18\\ \hline 
2.2 & 0 & 0 & 0.03 & 0.13 & 0.4 & 0.95 & 1.93 & 3.44 & 5.6 & 8.47 & 12.09 & 16.43 & 21.47\\ \hline 
2.3 & 0 & 0 & 0.04 & 0.16 & 0.48 & 1.14 & 2.27 & 4.02 & 6.47 & 9.7 & 13.69 & 18.44 & 23.85\\ \hline 
2.4 & 0 & 0 & 0.04 & 0.19 & 0.58 & 1.35 & 2.66 & 4.66 & 7.42 & 11.01 & 15.4 & 20.54 & 26.31\\ \hline 
2.5 & 0 & 0.01 & 0.05 & 0.23 & 0.68 & 1.58 & 3.09 & 5.35 & 8.45 & 12.42 & 17.2 & 22.72 & 28.85\\ \hline 
2.6 & 0 & 0.01 & 0.06 & 0.27 & 0.8 & 1.84 & 3.56 & 6.11 & 9.56 & 13.91 & 19.09 & 24.98 & 31.44\\ \hline 
2.7 & 0 & 0.01 & 0.08 & 0.32 & 0.94 & 2.13 & 4.08 & 6.93 & 10.74 & 15.48 & 21.05 & 27.31 & 34.07\\ \hline 
2.8 & 0 & 0.01 & 0.09 & 0.38 & 1.09 & 2.45 & 4.65 & 7.82 & 12 & 17.13 & 23.09 & 29.69 & 36.73\\ \hline 
2.9 & 0 & 0.01 & 0.11 & 0.44 & 1.26 & 2.8 & 5.26 & 8.76 & 13.32 & 18.85 & 25.18 & 32.11 & 39.4\\ \hline 
3 & 0 & 0.01 & 0.12 & 0.51 & 1.44 & 3.18 & 5.92 & 9.77 & 14.72 & 20.64 & 27.33 & 34.56 & 42.07\\ \hline 
3.1 & 0 & 0.02 & 0.15 & 0.59 & 1.65 & 3.59 & 6.62 & 10.83 & 16.17 & 22.48 & 29.53 & 37.03 & 44.72\\ \hline 
3.2 & 0 & 0.02 & 0.17 & 0.68 & 1.87 & 4.04 & 7.38 & 11.96 & 17.69 & 24.38 & 31.76 & 39.51 & 47.36\\ \hline 
3.3 & 0 & 0.02 & 0.19 & 0.78 & 2.11 & 4.52 & 8.18 & 13.14 & 19.27 & 26.33 & 34.01 & 41.99 & 49.96\\ \hline 
3.4 & 0 & 0.03 & 0.22 & 0.88 & 2.38 & 5.03 & 9.03 & 14.37 & 20.89 & 28.31 & 36.29 & 44.46 & 52.51\\ \hline 
3.5 & 0 & 0.03 & 0.25 & 1 & 2.66 & 5.58 & 9.92 & 15.66 & 22.57 & 30.33 & 38.57 & 46.91 & 55.02\\ 
\hline
\end{array}
$$

\pagebreak

\fbox{$z=7$}

$$
\begin{array}{|c||c|c|c|c|c|c|c|c|c|c|c|c|c|c|c|}
\hline
r \backslash q & 0.02 & 0.04 & 0.06 & 0.08 & 0.1 & 0.12 & 0.14 & 0.16 & 0.18 & 0.2 & 0.22 & 0.24 & 0.26\\ \hline \hline
0.1 & 0 & 0 & 0 & 0 & 0 & 0 & 0 & 0 & 0 & 0.01 & 0.02 & 0.05 & 0.09\\ \hline 
0.2 & 0 & 0 & 0 & 0 & 0 & 0 & 0 & 0 & 0.01 & 0.01 & 0.03 & 0.06 & 0.13\\ \hline 
0.3 & 0 & 0 & 0 & 0 & 0 & 0 & 0 & 0 & 0.01 & 0.02 & 0.05 & 0.09 & 0.18\\ \hline 
0.4 & 0 & 0 & 0 & 0 & 0 & 0 & 0 & 0.01 & 0.01 & 0.03 & 0.07 & 0.13 & 0.25\\ \hline 
0.5 & 0 & 0 & 0 & 0 & 0 & 0 & 0 & 0.01 & 0.02 & 0.05 & 0.1 & 0.19 & 0.35\\ \hline 
0.6 & 0 & 0 & 0 & 0 & 0 & 0 & 0.01 & 0.02 & 0.04 & 0.07 & 0.15 & 0.27 & 0.49\\ \hline 
0.7 & 0 & 0 & 0 & 0 & 0 & 0 & 0.01 & 0.02 & 0.05 & 0.11 & 0.21 & 0.39 & 0.68\\ \hline 
0.8 & 0 & 0 & 0 & 0 & 0 & 0.01 & 0.02 & 0.04 & 0.08 & 0.16 & 0.31 & 0.55 & 0.93\\ \hline 
0.9 & 0 & 0 & 0 & 0 & 0 & 0.01 & 0.02 & 0.06 & 0.12 & 0.24 & 0.43 & 0.75 & 1.25\\ \hline 
1 & 0 & 0 & 0 & 0 & 0 & 0.01 & 0.04 & 0.08 & 0.17 & 0.34 & 0.6 & 1.03 & 1.67\\ \hline 
1.1 & 0 & 0 & 0 & 0 & 0.01 & 0.02 & 0.05 & 0.12 & 0.25 & 0.47 & 0.83 & 1.38 & 2.2\\ \hline 
1.2 & 0 & 0 & 0 & 0 & 0.01 & 0.03 & 0.08 & 0.17 & 0.34 & 0.64 & 1.11 & 1.82 & 2.85\\ \hline 
1.3 & 0 & 0 & 0 & 0 & 0.01 & 0.04 & 0.11 & 0.24 & 0.47 & 0.86 & 1.47 & 2.36 & 3.63\\ \hline 
1.4 & 0 & 0 & 0 & 0 & 0.02 & 0.06 & 0.15 & 0.32 & 0.63 & 1.14 & 1.9 & 3.02 & 4.57\\ \hline 
1.5 & 0 & 0 & 0 & 0.01 & 0.03 & 0.08 & 0.2 & 0.43 & 0.83 & 1.47 & 2.43 & 3.8 & 5.66\\ \hline 
1.6 & 0 & 0 & 0 & 0.01 & 0.04 & 0.11 & 0.27 & 0.57 & 1.08 & 1.88 & 3.06 & 4.71 & 6.92\\ \hline 
1.7 & 0 & 0 & 0 & 0.01 & 0.05 & 0.15 & 0.35 & 0.74 & 1.38 & 2.37 & 3.8 & 5.77 & 8.35\\ \hline 
1.8 & 0 & 0 & 0 & 0.02 & 0.07 & 0.19 & 0.46 & 0.94 & 1.73 & 2.94 & 4.66 & 6.97 & 9.94\\ \hline 
1.9 & 0 & 0 & 0 & 0.02 & 0.09 & 0.25 & 0.58 & 1.18 & 2.15 & 3.6 & 5.63 & 8.31 & 11.71\\ \hline 
2 & 0 & 0 & 0.01 & 0.03 & 0.11 & 0.32 & 0.73 & 1.47 & 2.64 & 4.36 & 6.73 & 9.81 & 13.64\\ \hline 
2.1 & 0 & 0 & 0.01 & 0.04 & 0.14 & 0.4 & 0.91 & 1.8 & 3.2 & 5.22 & 7.95 & 11.44 & 15.72\\ \hline 
2.2 & 0 & 0 & 0.01 & 0.05 & 0.18 & 0.49 & 1.12 & 2.18 & 3.83 & 6.18 & 9.3 & 13.22 & 17.95\\ \hline 
2.3 & 0 & 0 & 0.01 & 0.06 & 0.22 & 0.61 & 1.36 & 2.62 & 4.55 & 7.24 & 10.77 & 15.14 & 20.31\\ \hline 
2.4 & 0 & 0 & 0.01 & 0.08 & 0.28 & 0.74 & 1.63 & 3.12 & 5.34 & 8.41 & 12.36 & 17.18 & 22.79\\ \hline 
2.5 & 0 & 0 & 0.02 & 0.1 & 0.34 & 0.9 & 1.95 & 3.68 & 6.22 & 9.68 & 14.07 & 19.34 & 25.38\\ \hline 
2.6 & 0 & 0 & 0.02 & 0.12 & 0.41 & 1.07 & 2.3 & 4.3 & 7.19 & 11.06 & 15.89 & 21.6 & 28.05\\ \hline 
2.7 & 0 & 0 & 0.03 & 0.14 & 0.49 & 1.27 & 2.7 & 4.98 & 8.24 & 12.53 & 17.82 & 23.96 & 30.8\\ \hline 
2.8 & 0 & 0 & 0.03 & 0.17 & 0.59 & 1.5 & 3.15 & 5.73 & 9.38 & 14.11 & 19.84 & 26.41 & 33.59\\ \hline 
2.9 & 0 & 0 & 0.04 & 0.21 & 0.7 & 1.75 & 3.64 & 6.55 & 10.6 & 15.77 & 21.95 & 28.92 & 36.43\\ \hline 
3 & 0 & 0 & 0.05 & 0.25 & 0.82 & 2.04 & 4.18 & 7.44 & 11.9 & 17.52 & 24.13 & 31.49 & 39.29\\ \hline 
3.1 & 0 & 0 & 0.06 & 0.29 & 0.95 & 2.35 & 4.76 & 8.39 & 13.29 & 19.36 & 26.39 & 34.1 & 42.15\\ \hline 
3.2 & 0 & 0.01 & 0.07 & 0.34 & 1.11 & 2.69 & 5.4 & 9.41 & 14.75 & 21.26 & 28.7 & 36.73 & 45\\ \hline 
3.3 & 0 & 0.01 & 0.08 & 0.4 & 1.28 & 3.07 & 6.09 & 10.5 & 16.28 & 23.24 & 31.06 & 39.39 & 47.83\\ \hline 
3.4 & 0 & 0.01 & 0.09 & 0.46 & 1.46 & 3.49 & 6.84 & 11.66 & 17.88 & 25.27 & 33.46 & 42.04 & 50.62\\ \hline 
3.5 & 0 & 0.01 & 0.11 & 0.54 & 1.67 & 3.93 & 7.63 & 12.88 & 19.55 & 27.36 & 35.88 & 44.69 & 53.36\\ 
\hline
\end{array}
$$

\pagebreak

\fbox{$z=8$}

$$
\begin{array}{|c||c|c|c|c|c|c|c|c|c|c|c|c|c|c|c|}
\hline
r \backslash q & 0.02 & 0.04 & 0.06 & 0.08 & 0.1 & 0.12 & 0.14 & 0.16 & 0.18 & 0.2 & 0.22 & 0.24 & 0.26\\ \hline \hline
0.1 & 0 & 0 & 0 & 0 & 0 & 0 & 0 & 0 & 0 & 0 & 0.01 & 0.01 & 0.03\\ \hline 
0.2 & 0 & 0 & 0 & 0 & 0 & 0 & 0 & 0 & 0 & 0 & 0.01 & 0.02 & 0.05\\ \hline 
0.3 & 0 & 0 & 0 & 0 & 0 & 0 & 0 & 0 & 0 & 0.01 & 0.02 & 0.03 & 0.07\\ \hline 
0.4 & 0 & 0 & 0 & 0 & 0 & 0 & 0 & 0 & 0 & 0.01 & 0.02 & 0.05 & 0.11\\ \hline 
0.5 & 0 & 0 & 0 & 0 & 0 & 0 & 0 & 0 & 0.01 & 0.02 & 0.04 & 0.08 & 0.16\\ \hline 
0.6 & 0 & 0 & 0 & 0 & 0 & 0 & 0 & 0 & 0.01 & 0.03 & 0.06 & 0.12 & 0.23\\ \hline 
0.7 & 0 & 0 & 0 & 0 & 0 & 0 & 0 & 0.01 & 0.02 & 0.04 & 0.09 & 0.18 & 0.33\\ \hline 
0.8 & 0 & 0 & 0 & 0 & 0 & 0 & 0 & 0.01 & 0.03 & 0.07 & 0.14 & 0.26 & 0.48\\ \hline 
0.9 & 0 & 0 & 0 & 0 & 0 & 0 & 0.01 & 0.02 & 0.05 & 0.1 & 0.2 & 0.38 & 0.68\\ \hline 
1 & 0 & 0 & 0 & 0 & 0 & 0 & 0.01 & 0.03 & 0.07 & 0.15 & 0.3 & 0.55 & 0.96\\ \hline 
1.1 & 0 & 0 & 0 & 0 & 0 & 0.01 & 0.02 & 0.05 & 0.11 & 0.23 & 0.43 & 0.77 & 1.32\\ \hline 
1.2 & 0 & 0 & 0 & 0 & 0 & 0.01 & 0.03 & 0.07 & 0.16 & 0.33 & 0.61 & 1.07 & 1.78\\ \hline 
1.3 & 0 & 0 & 0 & 0 & 0 & 0.01 & 0.04 & 0.11 & 0.23 & 0.46 & 0.84 & 1.45 & 2.36\\ \hline 
1.4 & 0 & 0 & 0 & 0 & 0.01 & 0.02 & 0.06 & 0.15 & 0.33 & 0.64 & 1.14 & 1.93 & 3.09\\ \hline 
1.5 & 0 & 0 & 0 & 0 & 0.01 & 0.03 & 0.09 & 0.21 & 0.45 & 0.86 & 1.52 & 2.52 & 3.96\\ \hline 
1.6 & 0 & 0 & 0 & 0 & 0.01 & 0.05 & 0.13 & 0.29 & 0.61 & 1.15 & 1.99 & 3.24 & 5\\ \hline 
1.7 & 0 & 0 & 0 & 0 & 0.02 & 0.06 & 0.17 & 0.4 & 0.81 & 1.5 & 2.56 & 4.1 & 6.22\\ \hline 
1.8 & 0 & 0 & 0 & 0.01 & 0.03 & 0.09 & 0.23 & 0.53 & 1.06 & 1.92 & 3.24 & 5.1 & 7.62\\ \hline 
1.9 & 0 & 0 & 0 & 0.01 & 0.04 & 0.12 & 0.31 & 0.69 & 1.36 & 2.43 & 4.03 & 6.26 & 9.21\\ \hline 
2 & 0 & 0 & 0 & 0.01 & 0.05 & 0.15 & 0.4 & 0.88 & 1.72 & 3.03 & 4.95 & 7.58 & 10.99\\ \hline 
2.1 & 0 & 0 & 0 & 0.01 & 0.06 & 0.2 & 0.51 & 1.12 & 2.14 & 3.73 & 6 & 9.06 & 12.94\\ \hline 
2.2 & 0 & 0 & 0 & 0.02 & 0.08 & 0.26 & 0.65 & 1.4 & 2.64 & 4.53 & 7.19 & 10.7 & 15.08\\ \hline 
2.3 & 0 & 0 & 0 & 0.02 & 0.11 & 0.33 & 0.82 & 1.72 & 3.22 & 5.44 & 8.51 & 12.49 & 17.38\\ \hline 
2.4 & 0 & 0 & 0 & 0.03 & 0.13 & 0.41 & 1.01 & 2.1 & 3.87 & 6.46 & 9.98 & 14.45 & 19.84\\ \hline 
2.5 & 0 & 0 & 0.01 & 0.04 & 0.17 & 0.51 & 1.24 & 2.54 & 4.61 & 7.6 & 11.57 & 16.54 & 22.43\\ \hline 
2.6 & 0 & 0 & 0.01 & 0.05 & 0.21 & 0.63 & 1.5 & 3.04 & 5.44 & 8.85 & 13.3 & 18.78 & 25.15\\ \hline 
2.7 & 0 & 0 & 0.01 & 0.06 & 0.26 & 0.77 & 1.8 & 3.6 & 6.36 & 10.21 & 15.16 & 21.14 & 27.97\\ \hline 
2.8 & 0 & 0 & 0.01 & 0.08 & 0.32 & 0.92 & 2.15 & 4.23 & 7.38 & 11.69 & 17.14 & 23.61 & 30.87\\ \hline 
2.9 & 0 & 0 & 0.01 & 0.1 & 0.39 & 1.11 & 2.53 & 4.93 & 8.49 & 13.27 & 19.23 & 26.17 & 33.84\\ \hline 
3 & 0 & 0 & 0.02 & 0.12 & 0.47 & 1.31 & 2.97 & 5.7 & 9.69 & 14.97 & 21.42 & 28.82 & 36.85\\ \hline 
3.1 & 0 & 0 & 0.02 & 0.14 & 0.56 & 1.55 & 3.45 & 6.54 & 10.98 & 16.76 & 23.71 & 31.54 & 39.89\\ \hline 
3.2 & 0 & 0 & 0.03 & 0.17 & 0.66 & 1.81 & 3.98 & 7.46 & 12.37 & 18.64 & 26.07 & 34.3 & 42.93\\ \hline 
3.3 & 0 & 0 & 0.03 & 0.21 & 0.78 & 2.11 & 4.57 & 8.45 & 13.84 & 20.62 & 28.51 & 37.1 & 45.96\\ \hline 
3.4 & 0 & 0 & 0.04 & 0.25 & 0.91 & 2.43 & 5.21 & 9.52 & 15.39 & 22.67 & 31 & 39.92 & 48.96\\ \hline 
3.5 & 0 & 0 & 0.05 & 0.29 & 1.06 & 2.79 & 5.91 & 10.66 & 17.03 & 24.8 & 33.53 & 42.74 & 51.92\\  
\hline
\end{array}
$$

\pagebreak

\fbox{$z=9$}

$$
\begin{array}{|c||c|c|c|c|c|c|c|c|c|c|c|c|c|c|c|}
\hline
r \backslash q & 0.02 & 0.04 & 0.06 & 0.08 & 0.1 & 0.12 & 0.14 & 0.16 & 0.18 & 0.2 & 0.22 & 0.24 & 0.26\\ \hline \hline
0.1 & 0 & 0 & 0 & 0 & 0 & 0 & 0 & 0 & 0 & 0 & 0 & 0 & 0.01\\ \hline
0.2 & 0 & 0 & 0 & 0 & 0 & 0 & 0 & 0 & 0 & 0 & 0 & 0.01 & 0.02\\ \hline
0.3 & 0 & 0 & 0 & 0 & 0 & 0 & 0 & 0 & 0 & 0 & 0.01 & 0.01 & 0.03\\ \hline
0.4 & 0 & 0 & 0 & 0 & 0 & 0 & 0 & 0 & 0 & 0 & 0.01 & 0.02 & 0.05\\ \hline
0.5 & 0 & 0 & 0 & 0 & 0 & 0 & 0 & 0 & 0 & 0.01 & 0.01 & 0.03 & 0.07\\ \hline
0.6 & 0 & 0 & 0 & 0 & 0 & 0 & 0 & 0 & 0 & 0.01 & 0.02 & 0.05 & 0.11\\ \hline
0.7 & 0 & 0 & 0 & 0 & 0 & 0 & 0 & 0 & 0.01 & 0.02 & 0.04 & 0.08 & 0.17\\ \hline
0.8 & 0 & 0 & 0 & 0 & 0 & 0 & 0 & 0 & 0.01 & 0.03 & 0.06 & 0.13 & 0.25\\ \hline
0.9 & 0 & 0 & 0 & 0 & 0 & 0 & 0 & 0.01 & 0.02 & 0.04 & 0.1 & 0.19 & 0.37\\ \hline
1 & 0 & 0 & 0 & 0 & 0 & 0 & 0 & 0.01 & 0.03 & 0.07 & 0.15 & 0.29 & 0.55\\ \hline
1.1 & 0 & 0 & 0 & 0 & 0 & 0 & 0.01 & 0.02 & 0.05 & 0.11 & 0.23 & 0.43 & 0.79\\ \hline
1.2 & 0 & 0 & 0 & 0 & 0 & 0 & 0.01 & 0.03 & 0.08 & 0.17 & 0.34 & 0.63 & 1.11\\ \hline
1.3 & 0 & 0 & 0 & 0 & 0 & 0.01 & 0.02 & 0.05 & 0.11 & 0.25 & 0.49 & 0.89 & 1.54\\ \hline
1.4 & 0 & 0 & 0 & 0 & 0 & 0.01 & 0.03 & 0.07 & 0.17 & 0.36 & 0.69 & 1.24 & 2.09\\ \hline
1.5 & 0 & 0 & 0 & 0 & 0 & 0.01 & 0.04 & 0.11 & 0.24 & 0.51 & 0.96 & 1.68 & 2.78\\ \hline
1.6 & 0 & 0 & 0 & 0 & 0 & 0.02 & 0.06 & 0.15 & 0.35 & 0.7 & 1.3 & 2.24 & 3.63\\ \hline
1.7 & 0 & 0 & 0 & 0 & 0.01 & 0.03 & 0.08 & 0.21 & 0.48 & 0.95 & 1.73 & 2.92 & 4.65\\ \hline
1.8 & 0 & 0 & 0 & 0 & 0.01 & 0.04 & 0.12 & 0.3 & 0.65 & 1.26 & 2.26 & 3.75 & 5.87\\ \hline
1.9 & 0 & 0 & 0 & 0 & 0.01 & 0.05 & 0.16 & 0.4 & 0.86 & 1.65 & 2.9 & 4.73 & 7.27\\ \hline
2 & 0 & 0 & 0 & 0 & 0.02 & 0.08 & 0.22 & 0.53 & 1.12 & 2.12 & 3.66 & 5.88 & 8.89\\ \hline
2.1 & 0 & 0 & 0 & 0.01 & 0.03 & 0.1 & 0.29 & 0.7 & 1.44 & 2.68 & 4.55 & 7.2 & 10.7\\ \hline
2.2 & 0 & 0 & 0 & 0.01 & 0.04 & 0.14 & 0.38 & 0.9 & 1.83 & 3.34 & 5.59 & 8.69 & 12.72\\ \hline
2.3 & 0 & 0 & 0 & 0.01 & 0.05 & 0.18 & 0.49 & 1.14 & 2.28 & 4.11 & 6.76 & 10.36 & 14.93\\ \hline
2.4 & 0 & 0 & 0 & 0.01 & 0.07 & 0.23 & 0.63 & 1.43 & 2.82 & 4.99 & 8.09 & 12.2 & 17.33\\ \hline
2.5 & 0 & 0 & 0 & 0.02 & 0.09 & 0.29 & 0.79 & 1.77 & 3.44 & 5.99 & 9.56 & 14.22 & 19.91\\ \hline
2.6 & 0 & 0 & 0 & 0.02 & 0.11 & 0.37 & 0.98 & 2.16 & 4.14 & 7.11 & 11.19 & 16.39 & 22.63\\ \hline
2.7 & 0 & 0 & 0 & 0.03 & 0.14 & 0.46 & 1.21 & 2.62 & 4.94 & 8.36 & 12.96 & 18.72 & 25.5\\ \hline
2.8 & 0 & 0 & 0 & 0.04 & 0.17 & 0.57 & 1.47 & 3.14 & 5.83 & 9.73 & 14.87 & 21.19 & 28.47\\ \hline
2.9 & 0 & 0 & 0.01 & 0.05 & 0.22 & 0.7 & 1.77 & 3.73 & 6.83 & 11.22 & 16.92 & 23.78 & 31.54\\ \hline
3 & 0 & 0 & 0.01 & 0.06 & 0.27 & 0.85 & 2.12 & 4.39 & 7.92 & 12.84 & 19.09 & 26.48 & 34.68\\ \hline
3.1 & 0 & 0 & 0.01 & 0.07 & 0.33 & 1.03 & 2.51 & 5.13 & 9.12 & 14.57 & 21.38 & 29.27 & 37.87\\ \hline
3.2 & 0 & 0 & 0.01 & 0.09 & 0.4 & 1.23 & 2.95 & 5.94 & 10.42 & 16.42 & 23.77 & 32.14 & 41.08\\ \hline
3.3 & 0 & 0 & 0.01 & 0.11 & 0.48 & 1.45 & 3.45 & 6.83 & 11.82 & 18.37 & 26.26 & 35.06 & 44.29\\ \hline
3.4 & 0 & 0 & 0.02 & 0.13 & 0.57 & 1.71 & 4 & 7.81 & 13.31 & 20.43 & 28.82 & 38.02 & 47.48\\ \hline
3.5 & 0 & 0 & 0.02 & 0.16 & 0.67 & 1.99 & 4.6 & 8.86 & 14.9 & 22.57 & 31.45 & 40.99 & 50.64\\
\hline
\end{array}
$$

\end{document}